\documentclass[prd,twocolumn,floatfix,amsmath,nofootinbib,amssymb,floatfix]{revtex4}
\usepackage{graphicx,color,dcolumn,booktabs,bm}
\usepackage{longtable,lscape}
\usepackage{pdfpages}
\usepackage{txfonts}
\usepackage{overpic}
\usepackage{amssymb}
\usepackage{makecell}
\usepackage{indentfirst}
\usepackage{feynmf}   
\usepackage{slashed}  
\usepackage{cases}
\usepackage{color}
\usepackage{multirow}
\usepackage{threeparttable}
\usepackage{epstopdf}
\usepackage{enumerate}
\usepackage{subfigure}
\usepackage{diagbox}
\usepackage{graphicx,color,dcolumn,booktabs,bm}
\usepackage[colorlinks,
            citecolor=blue,
            anchorcolor=red,
            menucolor=red,
            linkcolor=red,
            filecolor=red,
            runcolor=red,
            urlcolor=blue,
            frenchlinks=red]{hyperref}

\begin{document}
\title{Analysis of the form factors $B_c\rightarrow D^{(*)}$, $D_{s}^{(*)}$ and relevant nonleptonic decays}
\author{Bin Wu$^{1}$}
\author{Guo-Liang Yu$^{1,2}$}
\email{yuguoliang2011@163.com}
\author{Zhi-Gang Wang$^{1}$}
\email{zgwang@aliyun.com}
\author{Ze Zhou$^{1}$}
\author{Jie Lu$^{3}$}

\affiliation{$^1$ Department of Mathematics and Physics, North China
Electric Power University, Baoding 071003, China\\$^2$ Hebei Key Laboratory of Physics and Energy Technology, North China Electric Power University, Baoding 071000, China\\
$^3$ School of Physics, Southeast University, Nanjing 210094, China}
\date{\today}

\begin{abstract}
This article is devoted to calculating the form factors of $B_c \to D^{*}$, $B_c \to D$, $B_c \to D_s^{*}$ and $B_c \to D_s$ transitions in the framework of three-point QCD sum rules. At the QCD side, the contributions of $\langle\overline{q}q\rangle$, $\langle\overline{q}g_{s}\sigma Gq\rangle$, $\langle g_{s}^{2}G^{2}\rangle$, $\langle f^{3}G^{3}\rangle$ and $\langle\overline{q}q\rangle \langle g_{s}^{2}G^{2}\rangle$ are taken into account. With the obtained form factors, the decay widths and branching ratios of several two-body nonleptonic decay processes $B_c \to \eta_c D^{*}$, $\eta_c D$, $ J/\psi D^{*}$, $ J/\psi D$, $\eta_c D_s^{*}$, $\eta_c D_s$, $J/\psi D_s^{*}$ and $J/\psi D_s$ are predicted.  These results about  the form factors and decay properties of $B_c$ meson provide useful information for us to study the heavy-quark dynamics.
\end{abstract}

\pacs{13.25.Ft; 14.40.Lb}

\maketitle

\section{Introduction}\label{sec1}
The pseudoscalar meson $B_c$ which is composed of two heavy quarks with different flavours is an excellent laboratory to study the $B$ physics. Because each heavy quark in the $B_c$ meson can decay individually with the other acting as a spectator, $B_c$ is expected to have more rich decay channels than other $B$ mesons. Moreover, it was estimated that the inclusive production cross section of $B_c$ meson including its excited states at the LHC is at a level of $1ub$ for $\sqrt{14}$ TeV. This means that $O(10^9)$ $B_c$ mesons can be anticipated with 1 fb$^{-1}$\cite{Gao:2010zzc}. Similar viewpoint was also proposed in Ref. \cite{Chen:2018obq}. Thus, the abundant events in experiment have encouraged physicists to pay more attention to it.

$B_c$ is the lowest bound state consisted of $b$ and $c$ quarks, and lies below the threshold of decaying into the pair of heavy $B$ and $D$ mesons. Thus, pure electromagentic and strong decaying processes which are of flavor conservation are forbidden. $B_c$ can only decay according to weak interaction and is comparatively long-lived. There are three decay modes for $B_c$ meson, which are $c$-quark decaying channel with $b$ as a spectator, $b$-quark channel with $c$ as spectator and the $\bar{b}c$ annihilation channel $B_c^+ \to l^+\nu_{l}(c\bar{s}, u\bar{s})$, where $l=e, \nu, \tau$. The ratios of these processes are $~45\%$, $37\%$ and $18\%$ respectively\cite{Gershtein:1994jw}. However, only the former two decay processes were confirmed by experiment, such as $B_c \to J/\psi \pi$ and $B_c \to B_s^0 \pi$ decay channels\cite{LHCb:2012ihf, LHCb:2013xlg}.

The decay processes $B_c \to J/\psi D_s$ and $B_c \to J/\psi D_s^{*}$ were observed by LHCb experiment with high significance, and the following ratios of branching decays were measured\cite{LHCb:2013kwl},
\begin{align}
	\notag
	\frac{B(B_c \to J/\psi D_s)}{B(B_c \to J/\psi \pi)} = 2.90 \pm 0.57 (\mathrm{stat.}) \pm  0.24(\mathrm{syst.})\\
	\notag
	\frac{B(B_c \to J/\psi D_s^*)}{B(B_c \to J/\psi D_s)} = 2.37 \pm 0.56(\mathrm{stat.}) \pm 0.10(\mathrm{syst.})
\end{align}
The progresses made in experiment have motivated physicist to conduct more deeper studies on these nonleptonic decays about $B_c$ meson in theory, which can enhance our understanding about the heavy-quark dynamical behavior. The analysis about decay processes $B_c \to D^{(*)}$ and $B_c \to D_s^{(*)}$ can be carried out by different kinds of methods such as perturbative QCD (pQCD)\cite{Rui:2011qc, Rui:2012qq, Zou:2012sy, Wang:2014yia, Zhang:2024kjf}, QCD sum rules (QCDSR)\cite{Kiselev:2002vz, Azizi:2007jx, Azizi:2008vv, Azizi:2008vy,Wang:2024fwc}, Bauer-Stech-Wirbel (BSW) relativistic quark model\cite{Dhir:2008hh, Dhir:2008zz}, the covariant light-front quark model (CLFQM)\cite{Wang:2008xt,Chang:2019mmh,Li:2023wgq, Zhang:2023ypl, Li:2023mrj}, covariant confined quark model\cite{Dubnicka:2017job}, relativistic quark model\cite{Ivanov:2002un, Ivanov:2006ib}, light-cone QCD sum rules\cite{Huang:2008zg}(LCSR), etc. The QCDSR is a powerful non-perturbative method to study the properties of hadrons containing heavy quarks, and has made great achievements in predictions of the mass spectra, form factors, coupling constants and decay constants\cite{Colangelo:1992cx, Wang:2007fs, Aliev:2010ac, Wang:2012kw, Bashiry:2013waa, Peng:2019apl, Shi:2019hbf, Lu:2023gmd, Lu:2023lvu}.

Theoretically, the decay processes $B_c \to D^{(*)}$ and $B_c \to D_s^{(*)}$ are accompanied by $b \to c\bar c d$ and $b \to c\bar c s$ transitions, which can be described by effective Hamiltonian. The long distance dynamical behaviors are parameterized as weak form factors, which are complicated on account of the non-perturbative QCD effects in the bound hadron states. To study the nonleptonic decay processes of $B_c$ meson, we should know the values of the form factors by on-shell condition $q^{2}=M^{2}$ where $q$ is momentum transfer and $M$ is the mass of final state meson. In our previous work, form factors of $B_c \to \eta_c$ and $B_c \to J/\psi$ were calculated with the QCDSR\cite{Yu:2024utx}. As a continuation of that work, the form factors of $B_c \to D^{(*)}$ and $B_c \to D_s^{(*)}$ are systematically analyzed using the same method and the two-body nonleptonic decays of $B_c$ decaying to charmonium plus $D^{(*)}$ or $D_s^{(*)}$ meson are also studied.

The article is organized as follows. In section \ref{sec2}, we introduce in detail how to analyze the form factors in the framework of three-point QCDSR. In section \ref{sec3}, the numerical results of form factors are obtained and the values of form factors at $q^2=0$ are compared to those of other collaborations. In section \ref{sec4}, the decay widths and branching ratios of several decay channels including $B_c^- \to D^-_{s} \eta_c $, $D^-_{s} J/\psi $, $D_{s}^{*-} \eta_c $, $D_{s}^{*-} J/\psi $, and $B_c^- \to D^- \eta_c $, $D^- J/\psi $, $D^{*-} \eta_c $, $D^{*-} J/\psi $ are obtained with factorization approach. Finally, a brief conclusion is presented in section\ref{sec5}.

\section{Three-point QCDSR for the form factors}\label{sec2}

In the framework of QCDSR, the form factors are obtained by equating correlation functions in phenomenological and QCD sides, where correlation function is represented in hadronic and quark-gluon languages respectively. Thus, the first step to obtain form factor is to write the following three-point correlation function,
\begin{align}\label{correlator}
	\Pi (p,p') = {i^2}\int d^4 x d^4 ye^{ipx}e^{i(p - p')y}\left\langle 0 \right|T\{ {J_X}(x){J}(y)J_{B_c}^ + (0)\} \left| 0 \right\rangle
\end{align}
where $T$ is the time order operation. For the form factors of $B_c \to D^{(*)}$ and $B_c \to D_{s}^{(*)}$, $X$ denotes the meson $D^{(*)}$ or $D_s^{(*)}$. $p$ and $p'$ are the momentums of $B_c$ and $D^{(*)}/D_s^{(*)}$ mesons, $J_{B_c}$ and $J_X$ are interpolating currents which have the same quantum numbers with these mesons. $J(y)$ is transition current which is extracted from the low-energy effective Hamiltonian. These currents are written as follows,
\begin{align}\label{eq:2}
	\notag
	J_{B_c}(0) &= \bar c(0)i{\gamma _5}b(0)\\
	\notag
	J_D(x) &= \bar c(x)i{\gamma _5}d(x)\\
	\notag
	J_{D_s}(x) &= \bar c(x)i{\gamma _5}s(x)\\
	\notag
	{J^{D^*}_\mu}(x) &= \bar c(x){\gamma _\mu }d(x)\\
	\notag
	{J^{D_s^*}_\mu}(x) &= \bar c(x){\gamma _\mu }s(x)\\
	J(y) &= \bar q(y)\Gamma b(y)
\end{align}
where $q$ in the last equation denotes $d$ and $s$ quark for form factors of $B_c \to D^{(*)}$ and $B_c \to D_s^{(*)}$, respectively. $\Gamma = I, \gamma_\mu, \gamma_\mu \gamma_5, \sigma _{\mu \nu }\gamma_5$, which are correspond to scalar, vector, axial vector, tensor form factors, respectively.

\subsection{The phenomenological side}\label{sec2.1}
In the phenomenological side, a complete set of intermediate hadronic states with same quantum numbers as the current operators $J_{B_c}$ and $J_X$ are inserted into the correlation function. After the ground-state contributions being isolated, the correlation functions can be expressed as,
\begin{align}\label{hadronic correlator}
	\notag
	\Pi (p,p') &= \frac{\left\langle 0 \right|J_X(0)\left| X(p') \right\rangle \left\langle B_c(p) \right|J_{B_c}^ + ({\rm{0}})\left| {\rm{0}} \right\rangle }{(m_{B_c}^2 - {p^2})(m_X^2 - p{'^2})} \\
	&\quad\times \left\langle X(p') |J(0)| B_c(p) \right\rangle  + h.r.
\end{align}
where $h.r.$ denotes the contributions of excited and continuum states. The meson transition matrix elements are parameterized by various form factors,
\begin{widetext}
\begin{align}\label{meson transition}
	\notag
	\left\langle {P(p')} |\bar q b| B_c(p) \right\rangle  = &{f_S}({q^2})\\
	\notag
	\left\langle {P(p')} |\bar q{\gamma _\mu }b| B_c(p) \right\rangle  = &{f_ + }(q^2)\Bigg( p_\mu +p'_\mu - \frac{m_{B_c}^2 - m_P^2}{q^2}q_\mu  \Bigg) + f_0(q^2)\frac{m_{{B_c}}^2 - m_P^2}{q^2}q_\mu \\
	\notag
	\left\langle P(p') |\bar q\sigma _{\mu \nu }\gamma _5 b | B_c(p) \right\rangle  = & - \frac{2f_T(q^2)}{m_{B_c} + m_P}\varepsilon _{\mu \nu \alpha \beta }p^\alpha p'^\beta \\
	\notag
	\left\langle {V(p',\xi )} |\bar q{\gamma _\mu }b | B_c(p) \right\rangle  = & \frac{2V({q^2})}{m_{B_c} + m_V}\varepsilon _{\mu \nu \alpha \beta }{\xi ^{*\nu }}{p^\alpha }{p'^\beta }\\
	\notag
	\left\langle {V(p',\xi )} |\bar q\gamma _\mu \gamma _5 b | B_c(p) \right\rangle  = &i(m_{B_c} + m_V)\Bigg( \xi _\mu ^* - \frac{\xi ^* \cdot q}{q^2}q_\mu \Bigg) A_1(q^2) - i\frac{\xi ^* \cdot q}{m_{B_c} + m_V}\Bigg( p_{\mu}+p'_{\mu} - \frac{m_{B_c}^2 - m_V^2}{q^2}q_\mu \Bigg)A_2(q^2) \\
	&+ i(\xi ^* \cdot q)\frac{2m_V}{q^2}q_\mu A_0(q^2)
\end{align}
\end{widetext}
where $q=p-p'$, $P$ denotes the pesudoscalar meson $D$ and $D_s$, $V$ represents the vector meson $D^*$ and $D_s^*$, $\xi$ is polarization vector of relevant vector meson. Form factors at zero-recoil point($q^{2}=0$) satisfy the following relations\cite{Faustov:2019mqr},
\begin{align}
	\notag
	f_+(0) &= f_0(0)\\
	A_0(0) &= \frac{m_{B_c}+m_V}{2m_V}A_1(0)-\frac{m_{B_c}-m_V}{2m_V}A_2(0)
\end{align}
The meson vacuum matrix elements in Eq. (\ref{hadronic correlator}) can be parameterized as decay constants,
\begin{align}\label{vacuum meson}
	\notag
	\left\langle 0 |{J_P}(0)| {P(p')} \right\rangle  =& \frac{f_P m_P^2}{m_{1} + m_{2}}\\
	\left\langle 0 |J_\mu ^V(0)| V(p') \right\rangle  =& f_V m_V \xi _\mu
\end{align}
where $m_{1}$ and $m_{2}$ are the masses of quarks consisted in pesudoscalar meson. Replacing matrix elements in Eq. (\ref{hadronic correlator}) by the expressions of Eqs. (\ref{meson transition}) and (\ref{vacuum meson}), we can expand the correlation function into different tensor structures. Taking the vector form factors of $B_c \to D$ transition as an example,
\begin{align}\label{hadronic structure}
	\notag
	\Pi _\mu ^{phy}(p,p') &= \frac{B\big[(1 - A)f_ + (q^2) + Af_0(q^2)\big]}{(m_D^2 - p'^2)(m_{B_c}^2 - p^2)}p_\mu  \\
	&+ \frac{B\big[(1 + A)f_+ (q^2) - A f_0(q^2)\big]}{(m_D^2 - p'^2)(m_{B_c}^2 - p^2)}p'_\mu
\end{align}
with,
\begin{align}
	\notag
	A &= \frac{m_{B_c}^2 - m_D^2}{q^2}\\
	B &= \frac{f_D m_D^2}{m_d + m_c}\times\frac{f_{B_c}m_{B_c}^2}{m_c + m_b}
\end{align}
In the QCD side, the correlation function will have same tensor structures as in phenomenological side. After equating both of these two sides with the same tensor structure, the form factors $f_+$ and $f_0$ can be expressed as linear combination of coefficients of the tensor structures $p_\mu$ and $p'_\mu$. We can also obtain the other form factors according to similar processes.

\subsection{The QCD side}\label{sec2.2}
\begin{figure}
	\centering
	\includegraphics[width=6cm, trim=0 0 0 0, clip]{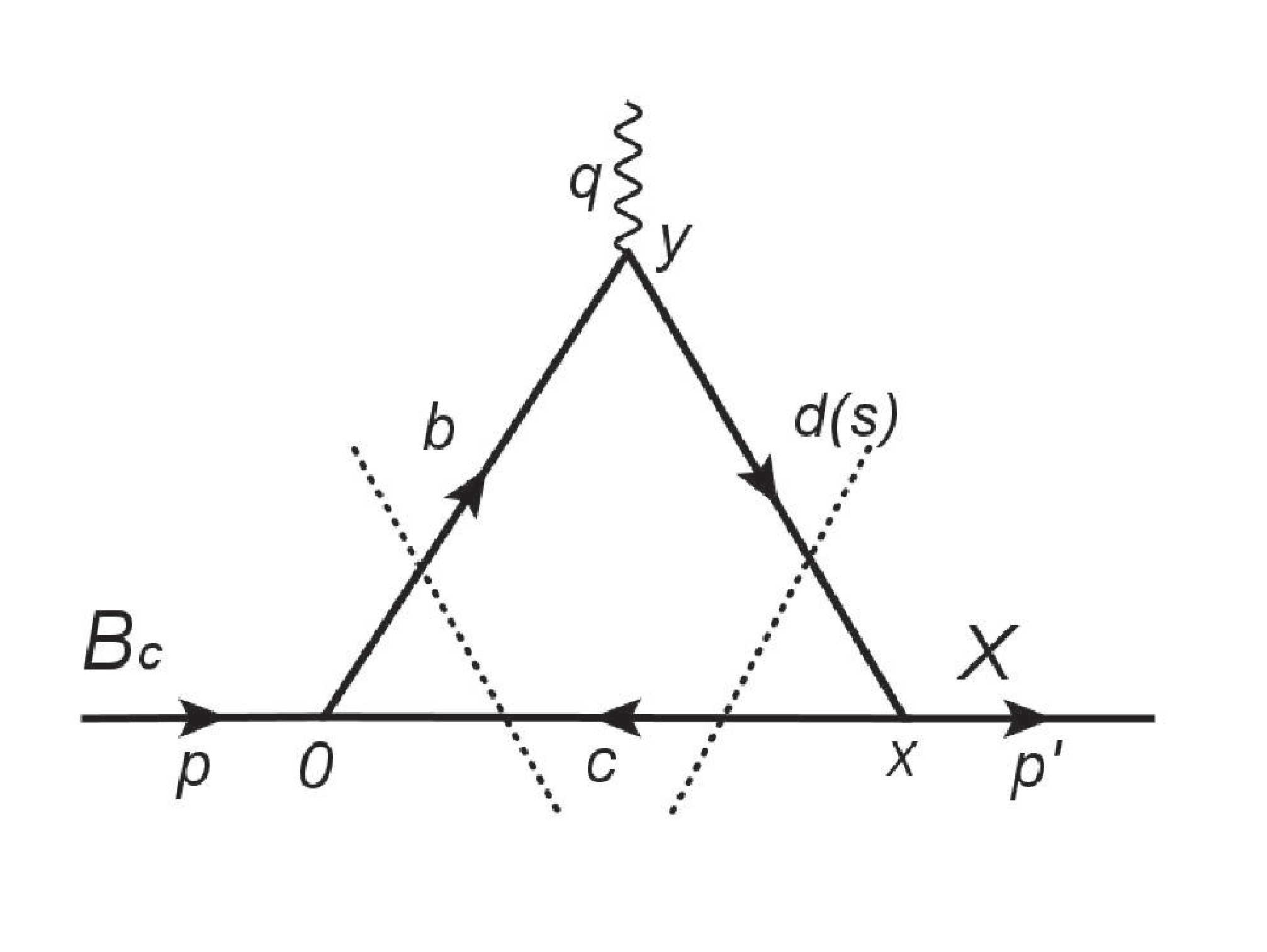}
	\caption{The Feynman diagram for the perturbative part. The dashed lines denote the Cutkosky's cuts.}
	\label{cutting rule}
\end{figure}

\begin{figure*}[htbp]
	\centering
	\includegraphics[width=16cm]{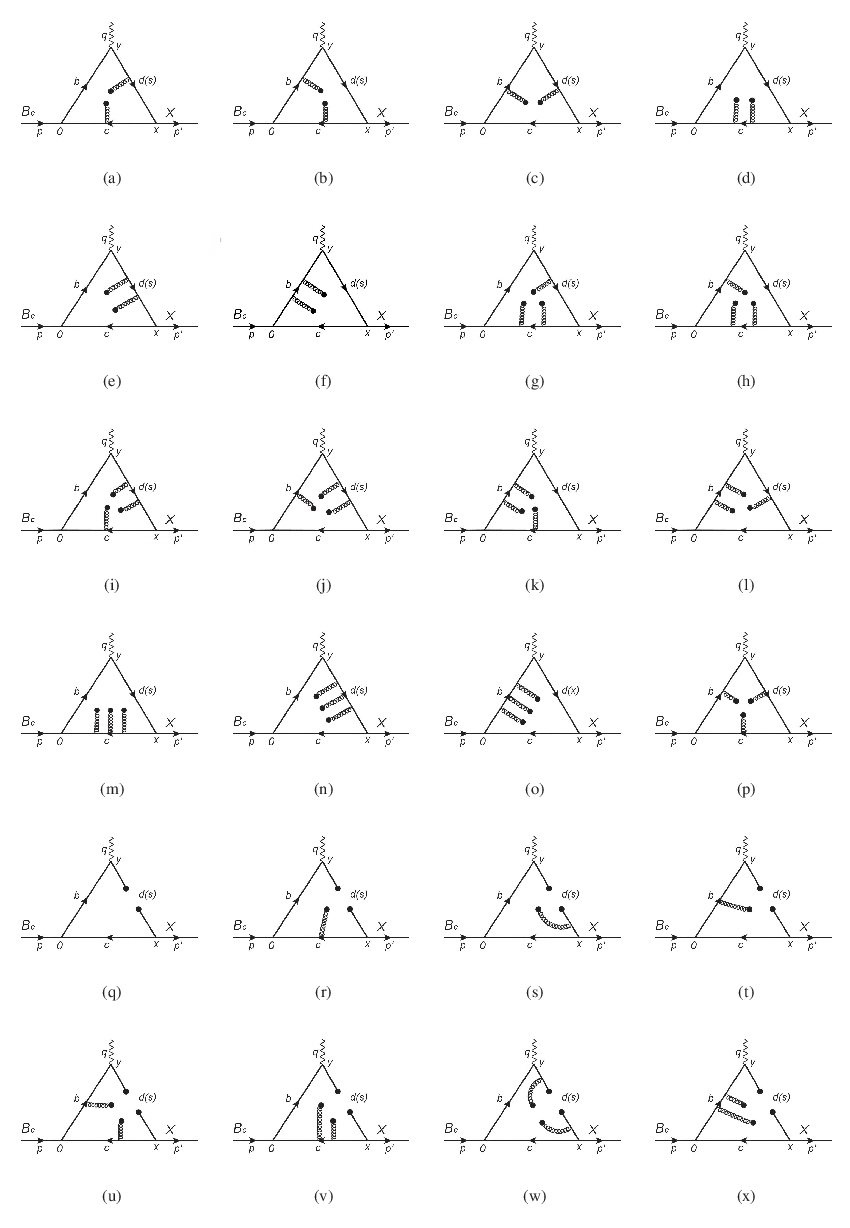}
	\caption{Feynman diagrams of vacuum condensates. Contributions from  quark condensate (q) and quark-gluon mixing condensates (r)-(x) are zero in the calculation.}
	\label{Feynman}
\end{figure*}

At quark level, the quark fields in the correlation function Eq. (\ref{correlator}) are contracted by using Wick's theorem. Thus, the correlation functions in the QCD side for $B_c \to D$, $B_c \to D^*$, $B_c \to D_s$ and $B_c \to D_s^*$ processes can be expressed as,
\begin{align}\label{QCDcorrelator}
	\notag
	&\Pi^{B_c \to D} (p,p') =  - \int {d^4 x} d^4y e^{ip'x} e^{i(p - p')y}\\
	\notag
	&\quad \times \left\langle 0 \big|Tr\Big[ C^{lm}(-x){\gamma _5}D^{mn}(x - y) B^{nl}(y)\gamma _5 \Big] \big| 0 \right\rangle \\
	\notag
	&\Pi_\mu^{B_c \to D} (p,p') =  - \int {d^4 x} d^4y e^{ip'x} e^{i(p - p')y}\\
	\notag
	&\quad \times \left\langle 0 \big|Tr\Big[ C^{lm}(-x){\gamma _5}D^{mn}(x - y)\gamma _\mu B^{nl}(y)\gamma _5 \Big] \big| 0 \right\rangle \\
	\notag
	&\Pi_{\mu\nu}^{B_c \to D} (p,p') =  - \int {d^4 x} d^4y e^{ip'x} e^{i(p - p')y}\\
	&\quad \times \left\langle 0 \big|Tr\Big[ C^{lm}(-x){\gamma _5}D^{mn}(x - y)\sigma_{\mu\nu} \gamma_5 B^{nl}(y)\gamma _5 \Big] \big| 0 \right\rangle
\end{align}
\begin{align}
	\notag
	&\Pi_\mu^{B_c \to D^*} (p,p') =  - \int {d^4 x} d^4y e^{ip'x} e^{i(p - p')y}\\
	\notag
	&\quad \times \left\langle 0 \big|Tr\Big[ C^{lm}(-x){\gamma _5}D^{mn}(x - y)\gamma _\mu B^{nl}(y)\gamma _5 \Big] \big| 0 \right\rangle \\
	\notag
	&\Pi_\mu^{B_c \to D^*} (p,p') =  - \int {d^4 x} d^4y e^{ip'x} e^{i(p - p')y}\\
	&\quad \times \left\langle 0 \big|Tr\Big[ C^{lm}(-x){\gamma _5}D^{mn}(x - y)\gamma _\mu \gamma_5 B^{nl}(y)\gamma _5 \Big] \big| 0 \right\rangle
\end{align}
\begin{align}
	\notag
	&\Pi^{B_c \to D_s} (p,p') =  - \int {d^4 x} d^4y e^{ip'x} e^{i(p - p')y}\\
	\notag
	&\quad \times \left\langle 0 \big|Tr\Big[ C^{lm}(-x){\gamma _5}S^{mn}(x - y) B^{nl}(y)\gamma _5 \Big] \big| 0 \right\rangle \\
	\notag
	&\Pi_\mu^{B_c \to D_s} (p,p') =  - \int {d^4 x} d^4y e^{ip'x} e^{i(p - p')y}\\
	\notag
	&\quad \times \left\langle 0 \big|Tr\Big[ C^{lm}(-x){\gamma _5}S^{mn}(x - y)\gamma _\mu B^{nl}(y)\gamma _5 \Big] \big| 0 \right\rangle \\
	\notag
	&\Pi_{\mu\nu}^{B_c \to D_s} (p,p') =  - \int {d^4 x} d^4y e^{ip'x} e^{i(p - p')y}\\
	&\quad \times \left\langle 0 \big|Tr\Big[ C^{lm}(-x){\gamma _5}S^{mn}(x - y)\sigma_{\mu\nu} \gamma_5 B^{nl}(y)\gamma _5 \Big] \big| 0 \right\rangle
\end{align}
\begin{align}
	\notag
	&\Pi_\mu^{B_c \to D_s^*} (p,p') =  - \int {d^4 x} d^4y e^{ip'x} e^{i(p - p')y}\\
	\notag
	&\quad \times \left\langle 0 \big|Tr\Big[ C^{lm}(-x){\gamma _5}S^{mn}(x - y)\gamma _\mu B^{nl}(y)\gamma _5 \Big] \big| 0 \right\rangle \\
	\notag
	&\Pi_\mu^{B_c \to D_s^*} (p,p') =  - \int {d^4 x} d^4y e^{ip'x} e^{i(p - p')y}\\
	&\quad \times \left\langle 0 \big|Tr\Big[ C^{lm}(-x){\gamma _5}S^{mn}(x - y)\gamma _\mu \gamma_5 B^{nl}(y)\gamma _5 \Big] \big| 0 \right\rangle
\end{align}
where $D^{ij}$, $S^{ij}$, $C^{ij}$and $B^{ij}$ are the full propagators of $d$, $s$, $c$ and $b$ quarks. These propagators can be written as follows\cite{Wang:2014yza},
\begin{align}
		\notag
& q^{ij}(x) = \frac{i\delta^{ij}\slashed{x}}{2\pi ^2x^4} - \frac{\delta^{ij}m_q}{4\pi ^2x^4} - \frac{\delta ^{ij}\left\langle  \bar qq \right\rangle}{12} + \frac{i\delta^{ij}\slashed{x} {m_q}\left\langle \bar qq \right\rangle } {48} \\
\notag
&- \frac{\delta ^{ij}{x^2}\left\langle {\bar q{g_s}\sigma Gq} \right\rangle }{192} + \frac{i\delta ^{ij}x^2\slashed{x}m_q\left\langle  {\bar q{g_s}\sigma Gq} \right\rangle }{1152} \\
\notag
&- \frac{ig_sG_{\alpha \beta }^at_{ij}^a(\slashed{x} {\sigma ^{\alpha \beta }} + \sigma ^{\alpha \beta }\slashed{x})}{32{\pi ^2}{x^2}}- \frac{i\delta^{ij}x^2\slashed{x} g_s^2{{\left\langle  {\bar qq} \right\rangle }^2}}{7776} \\
		&- \frac{\delta^{ij}{x^4}\left\langle  {\bar qq} \right\rangle \left\langle  {g_s^2{G^2}} \right\rangle }{27648} - \frac{\left\langle  {{{\bar q}^j}{\sigma ^{\mu \nu }}{q^i}} \right\rangle \sigma _{\mu \nu }}{8} + ...\\
		\notag
& Q^{ij}(x) = \frac{i}{(2\pi)^4}\int d^4 k e^{-ik \cdot x} \Bigg\{ \frac{\delta^{ij}}{\slashed k - {m_Q} } \\
\notag
&- \frac{g_s G_{\alpha \beta }^nt_{ij}^n}{4}\frac{\sigma ^{\alpha \beta }(\slashed k + m_Q) + (\slashed k + m_Q)\sigma ^{\alpha \beta }}{(k^2 - m_Q^2)^2 } \\
\notag
&+ \frac{g_s D_\alpha G_{\beta \lambda }^nt_{ij}^n(f^{\lambda \beta \alpha} + f^{\lambda \alpha \beta})}{3(k^2 - m_Q^2)^4 } \\ \notag
		& - \frac{g_s^2 (t^a t^b)_ij G_{\alpha \beta }^a G_{\mu \nu }^b(f^{\alpha \beta \mu \nu } + f^{\alpha \mu \beta \nu } + f^{\alpha \mu \nu \beta })}{4(k^2 - m_Q^2)^5 }+ \\
\notag
&\frac{i{\delta^{ij}}\left\langle g_s^3{G^3} \right\rangle }{48}\frac{(\slashed k + m_Q)\Big[\slashed k(k^2 - 3m_Q^2) + 2m_Q(2k^2 - m_Q^2)\Big](\slashed k + m_Q)}{({k^2} - m_Q^2)^6 } \\
&+  \ldots \Bigg\}
\end{align}
where $q^{ij}$ and $Q^{ij}$ denote light and heavy quark full propagators, respectively, $i$ and  $j$ are color indices, ${\sigma _{\alpha \beta }} = i[{\gamma _\alpha },{\gamma _\beta }]/2$, ${D_\alpha } = {\partial _\alpha } - i{g_s}G_\alpha ^n{t^n}$, $G_\alpha^n$ is the gluon field, ${t^n} = {\lambda ^n}/2$, and ${\lambda ^n}$ is the Gell-Mann matrix, ${f^{\lambda \beta \alpha }}$ and ${f^{\alpha \beta \mu \nu }}$ are defined as,
\begin{align}
	\notag
	{f^{\lambda \alpha \beta }} = &(\slashed k + {m_Q}){\gamma ^\lambda }(\slashed k + {m_Q}){\gamma ^a}(\slashed k + {m_Q}) {\gamma ^\beta }(\slashed k + {m_Q})\\
	{f^{\alpha \beta \mu \nu }} = &(\slashed k + {m_Q}){\gamma ^\alpha }(\slashed k + {m_Q}){\gamma ^\beta }(\slashed k + {m_Q}) {\gamma ^\mu }(\slashed k + {m_Q}){\gamma ^\nu }(\slashed k + {m_Q})
\end{align}
Performing the operator product expansion (OPE), the correlation functions are represented in different tensor structures in the QCD side as same as that in the phenomenological side. For the vector form factor $B_c \to D$ as an example, its correlation function is written as the following form,
\begin{align}
	\Pi _\mu ^{\rm{OPE}} = F_1(q^2)p_\mu  + F_2(q^2)p'_\mu
\end{align}
Here, $F_i(q^2)$ is called invariant amplitude which is a function of transfer momentum squared. For each Dirac structure, the invariant amplitude can be expressed as the spectra density $\rho(s,u,q^2)$ according to dispersion integral,
\begin{align}
	F_i(q^2) =  - \frac{1}{4\pi ^2}\int\limits_{s_{min}}^\infty  {\int\limits_{u_{min}}^\infty} dsdu \frac{\rho _i(s,u,q^2)}{(s - {p^2})(u - p'^2)}
\end{align}
where $s_{min}$ and $u_{min}$ are kinematic limits with their values to be $s_{min}=(m_c+m_b)^2$ and $u_{min}=(m_{d}/m_{s}+m_c)^2$ respectively. $\rho_i(s,u,q^2)$ is the QCD spectral density where $s=p^2$, $u=p'^2$. The spectral density is obtained from the imaginary part of correlation function and it originates from the contributions of perturbative and non-perturbative parts.
\begin{align}
	\rho _i = \rho _i^{pert} + \rho _i^{non-pert}
\end{align}
According to the Cutkosky's rule\cite{Cutkosky:1960sp, Wang:2007ys}, the spectral density of perturbative part can be obtained by putting all quark lines on shell  (Fig. \ref{cutting rule}). In this process, the following condition should be satisfied,
\begin{align}
	- 1 \le \frac{2s(m_d^2-m_c^2+u) - (m_b^2 - m_c^2 + u - {q^2})(s + u - {q^2})}{\sqrt {\left[ (m_b^2 - m_c^2 + u - {q^2})^2 - 4 s m_d^2 \right]\lambda (s,u,{q^2})} } \le 1
\end{align}
where $\lambda$ function has the following expression,
\begin{align}
	\lambda (a,b,c) = {a^2} + {b^2} + {c^2} - 2(ab + bc + ac)
\end{align}
The non-perturbative contribution is reflected in several vacuum condensates, including the quark condensate, two-gluon condensate, quark-gluon mixing condensate, three-gluon condensate and so on. These vacuum condensates are illustrated in Fig. \ref{Feynman}.
After lengthy and tedious derivation, we find that the contributions from quark condensate and quark-gluon mixing condensate only depend on $p^2$ or $p'^2$. Thus, these contributions are zero after performing the double Borel transformation with respect to both $p^2$ and $p'^2$. The spectral densities of gluon condensates can be obtained by similar procedure as that of perturbative part. To   obtain its spectral density by Cutkosky's rule, the following equation is used to reduce the power of the quark propagator,
\begin{align}
	\notag
	&\int d^4 k \frac{1}{[(k - p')^2 - m_c^2]^\alpha (k^2 - m_{d(s)}^2)^\beta [(k + p - p')^2 - m_b^2]^\gamma } \\
	\notag
	&= \frac{1}{(\alpha  - 1)!(\beta  - 1)!(\gamma  - 1)!}\frac{\partial ^{\alpha  - 1}}{\partial m_c^{\alpha  - 1}}\frac{\partial ^{\beta  - 1}}{\partial m_{d(s)}^{\beta - 1}}\frac{\partial ^{\gamma  - 1}}{\partial m_b^{\gamma - 1}}\\
	&\times \int d^4 k \frac{1}{[(k - p')^2 - m_c^2](k^2 - m_{d(s)}^2)[(k + p - p')^2 - m_b^2]}
\end{align}

\subsection{QCD sum rules for form factors}
After the spectral densities have been calculated, the sum rules for form factors can be obtained by matching the phenomenological side and the QCD side. To eliminate the contributions from excited and continuum states, the threshold parameters $s_{0}$ and $u_{0}$ are introduced, and the double Borel transformation is preformed to both the phenomenological side and the QCD side. Taking also the vector form factors of $B_c \to D$ transition as an example, the sum rules can be expressed as,
\begin{align}\label{sum rule}
	\notag
	&\bigg\{B\Big[(1 - A)f_+(Q^2) + Af_0(Q^2)\Big]p_\mu \\
	\notag
	& + B\Big[(1 + A)f_+(Q^2) - Af_0(Q^2)\Big]p'_\mu\bigg\}\times\exp \Big( -\frac{m_{{B_c}}^2}{M^2} - \frac{m_D^2}{kM^2} \Big) \\
	\notag
	&=  - \frac{1}{4 \pi^2}\int\limits_{s_{\min}}^{s_0} \int\limits_{u_{\min }}^{u_0} dsdu \Big[\rho _1(s,u,Q^2)p_\mu + \rho _2(s,u,Q^2)p'_\mu \Big] \\
	& \times \exp \Big(  - \frac{s}{M^2} - \frac{u}{kM^2} \Big)
\end{align}
where substitutions of $p^{2}\rightarrow -P^{2}$, $p'^{2}\rightarrow -P'^{2}$ and $q^{2}\rightarrow -Q^{2}$ are conducted, the threshold parameters $s_0$ and $u_0$ serve as the upper limits of integral. After performing double Borel transformation with respect to $P^{2}$ and $P'^{2}$, there are two Borel parameters $M^2$ and $M^{\prime2}$. We then take $M^{\prime2}=kM^2$ with the factor $k=m_{X}^2/m_{B_c}^2$\cite{Bracco:2011pg}. The double Borel transformation can enhance the contribution of ground state while suppress those of excited and continuum states. It is noted that the physical quantities, such as the form factor, extracted from QCD sum rules should be independent of Borel parameters. That is to say, we should find a plateau where the form factors are stable and convergent. This plateau is commonly called the Borel Window or Borel platform.

\section{Numerical results of the form factors}\label{sec3}
The masses of mesons used in this work are taken from the Particle Date Group (PDG)\cite{ParticleDataGroup:2024cfk}. The masses of quarks are energy-scale dependent and can be expressed as the following renormalization group equation,
\begin{align}
	\notag
	{m_{q}}(\mu ) &= m_{q}(m_{q})\Big[\frac{\alpha_s(\mu )}{\alpha_s(m_{q})}\Big]^{\frac{12}{33 - 2n_f}}\\
	\notag
	{m_{Q}}(\mu ) &= m_{Q}(m_{Q})\Big[\frac{\alpha_s(\mu )}{\alpha_s(m_{Q})}\Big]^{\frac{12}{33 - 2n_f}}\\
	\alpha_s(\mu) &= \frac{1}{b_0t}\Big[1 - \frac{b_1}{b_0^2}\frac{\log t}{t}
	+ \frac{b_1^2(\log^2 t - \log t - 1) + {b_0}{b_2}}{b_0^4 t^2}\Big]
\end{align}
where $t = \log ({\mu ^2}/{\Lambda _{QCD}^2})$, ${b_0} = ({33 - 2 n_f})/{12\pi}$, ${b_1} = ({153 - 19 n_f})/{24 \pi^2}$, ${b_2} = ({2857 - \frac{5033}{9}{n_f} + \frac{325}{27}n_f^2})/{128 \pi^3}$. The $\overline{\mathrm{MS}}$ masses are also taken from the PDG with $m_{c}(m_{c})=1.275\pm0.025$ GeV and $m_{b}(m_{b})=4.18\pm0.03$ GeV. $\Lambda _{QCD} = 213$ MeV for the flavors $n_f = 5$ and the energy-scales are uniformly determined to be $2$ GeV, which are also adopted in our previous work\cite{Yu:2024utx}. The decay constants of mesons are taken from Refs. \cite{Narison:2020guz} and \cite{Wang:2015mxa}, where these hadronic parameters are uniformly obtained by the QCDSR. The vacuum condensates are taken as standard values from Refs. \cite{Narison:2010cg, Narison:2011xe, Narison:2011rn}. Threshold parameters $s_0$ and $u_0$ are used to eliminate the contributions of excited and continuum states. Generally, their values are taken to be $s_0=(m_{B_c}+\Delta)^2$ and $u_0=(m_X+\Delta)^2$, where $X$ represents the final meson $D^{(*)}$ or $D_s^{(*)}$. Theoretically, the value of $\Delta$ should be larger than the width of ground state and be smaller than the distance between the ground state and the first excitation. In this work, $\Delta$ is chosen to be $0.4$, $0.5$ and $0.6$ GeV for the lowest, central and the highest values of form factors. All of the values of parameters used in this work are listed in Tab. \ref{parameters}.
\begin{table}[b]
	\begin{ruledtabular}
		\caption{Values of parameters used in this work, the values with no reference are mentioned in the text.}
		\begin{tabular}{c c c c}
			Parameters & Values(GeV) & Parameters & Values \\
			\hline
			$m_{B_c}$ & 6.27 & $f_{B_c}$ & 0.371 GeV\cite{Narison:2020guz} \\
			$m_{D}$ & 1.87 & $f_{D}$ & 0.208 GeV\cite{Wang:2015mxa} \\
			$m_{D^*}$ & 2.01 & $f_{D^*}$ & 0.263 GeV\cite{Wang:2015mxa} \\
			$m_{D_s}$ & 1.97 & $f_{D_s}$ & 0.240 GeV\cite{Wang:2015mxa} \\
			$m_{D^*_s}$ & 2.11 & $f_{D^*_s}$ & 0.308 GeV\cite{Wang:2015mxa} \\
			$m_{\eta_c}$ & 2.98 & $f_{\eta_c}$ & 0.387 GeV\cite{Becirevic:2013bsa} \\
			$m_{J/\psi}$ & 3.10 & $f_{J/\psi}$ & 0.418 GeV\cite{Becirevic:2013bsa} \\
			$m_s(\mu=2 \mathrm{GeV})$ & 0.095 & $\left\langle {g_s^2{G^2}} \right\rangle$ & $(0.88 \pm 0.15)$ GeV$^4$ \\
			$m_c(\mu=2 \mathrm{GeV})$ & 1.16 & $\left\langle {f{G^3}} \right\rangle $ & $(8.8 \pm 5.5)$ GeV$^2 \left\langle {g_s^2{G^2}} \right\rangle$ \\
			$m_b(\mu=2 \mathrm{GeV})$ & 4.76 & $V_{cb}$ & 0.041\cite{ParticleDataGroup:2024cfk} \\
			$V_{cu/d}$ & 0.221\cite{ParticleDataGroup:2024cfk} & $V_{cs}$ & 0.975\cite{ParticleDataGroup:2024cfk} \\
			$a_1$ & 1.07\cite{Zhang:2023ypl} & $a_2$ & 0.234 \cite{Zhang:2023ypl}
			\label{parameters}
		\end{tabular}
	\end{ruledtabular}
\end{table}

The Borel parameter $M^2$ is determined according to two conditions, which are the pole dominance($\geq40\%$) and convergence of OPE. The pole contribution is defined as follow\cite{Bracco:2011pg},
\begin{align}
	\mathrm{pole} = \frac{\Pi _\mathrm{pole}(M^2)}{\Pi _\mathrm{pole}(M^2)+\Pi _\mathrm{cout}(M^2)}
\end{align}
with,
\begin{align}
	\notag
	\Pi _\mathrm{pole}(M^2) &= - \frac{1}{4 \pi ^2} \int\limits_{s_{min}}^{s_0} \int\limits_{u_{min}}^{u_0} \rho ^{QCD}(s,u,Q^2) e^{ - \frac{s}{M^2} - \frac{u}{kM^2}}dsdu \\
	\Pi _\mathrm{cout}(M^2) &= - \frac{1}{4 \pi ^2} \int\limits_{s_0}^\infty \int\limits_{u_0}^\infty \rho ^{QCD}(s,u,Q^2)e^{ - \frac{s}{M^2} - \frac{u}{k M^2}}dsdu
\end{align}
\begin{figure}[htbp]
	\centering
	\includegraphics[width=8.5cm]{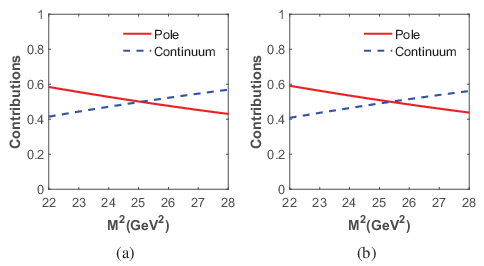}
	\caption{The pole contributions of vector form factors of $B_c \to D$ transition. Subgraph (a) and (b) correspond to $f_+^{B_c \to D}$ and $f_0^{B_c \to D}$, respectively.}
	\label{PC example}
\end{figure}
\begin{figure}[htbp]
	\centering
	\includegraphics[width=8.5cm]{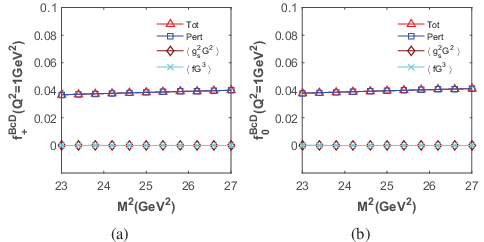}
	\caption{The contributions of perturbative part and different vacuum condensate terms with variation of the Borel parameter.}
	\label{BW example}
\end{figure}

Still taking vector form factor of $B_c \to D$ as an example, we introduce how the Borel parameters are determined. Fixing $Q^2 = 1$ GeV$^2$, we plot the variation of pole contribution with Borel parameter $M^2$ in Fig. \ref{PC example}. It is shown that the pole contribution decreases with increase of Borel parameter $M^2$. When Borel parameter is smaller than $28$ GeV$^2$, pole contribution is higher than $40\%$. To find the Borel platform where the condition of OPE convergence is satisfied and the results have good stability and convergence, the contributions of perturbative part and all vacuum condensate terms are plotted in Fig. \ref{BW example}. It is obviously shown that the contribution of perturbative term is dominant in the region $23$ GeV$^2$$\leq M^2\leq27$ GeV$^2$, while the contributions from vacuum condensate terms are much less than the former. That is to say, the OPE convergence is well satisfied. According to these above analyses, the Borel platform is determined to be $23$$-27$ GeV$^2$ where the conditions of the pole dominance($\geq40\%$) and convergence of OPE are all satisfied. The values of Borel parameters and pole contributions for different form factors are all listed in Tab. \ref{region}.
\begin{table}[htbp]
	\begin{ruledtabular}
		\caption{The Borel platform and pole contribution for different form factors}
		\begin{tabular}{c >{\centering}c c c }
			Modes & Form factors & Borel platforms & Pole contributions($\%$) \\
			\hline
			\multirow{4}{*}{$B_c \to D$}
			& $f_s$ & 20 $\sim$ 24 & 51.27\\
			& $f_+$ & 23 $\sim$ 27 & 50.20\\
			& $f_0$ & 23 $\sim$ 27 & 50.95\\
			& $f_T$ & 30 $\sim$ 34 & 50.02\\
			\hline
			\multirow{4}{*}{$B_c \to  D^*$}
			& $V$ & 33 $\sim$ 37 & 50.12\\
			& $A_0$ & 25 $\sim$ 29 & 51.78\\
			& $A_1$ & 22 $\sim$ 26 & 50.48\\
			& $A_2$ & 19 $\sim$ 23 & 51.34\\
			\hline
			\multirow{4}{*}{$B_c \to D_s$}
			& $f_s$ & 21 $\sim$ 25 & 51.58\\
			& $f_+$ & 24 $\sim$ 28 & 50.91\\
			& $f_0$ & 24 $\sim$ 28 & 51.65\\
			& $f_T$ & 31 $\sim$ 35 & 50.70\\
			\hline
			\multirow{4}{*}{$B_c \to  D_s^*$}
			& $V$ & 33 $\sim$ 37 & 51.03\\
			& $A_0$ & 26 $\sim$ 30 & 50.91\\
			& $A_1$ & 22 $\sim$ 26 & 52.59\\
			& $A_2$ & 20 $\sim$ 24 & 50.13
			\label{region}
		\end{tabular}
	\end{ruledtabular}
\end{table}
\begin{table}[htbp]
	\begin{ruledtabular}
		\caption{Fitting parameters of the $z$-series parameterized approach.}
		\begin{tabular}{cccccc}
			& Modes & Form factors & $b_0$ & $b_1$ & $b_2$ \\
			\hline
			& \multirow{4}{*}{$B_c \to D$} & $f_s$ & 0.54 & -5.9 & 14  \\
			&& $f_+$ & 0.093 & -1.4 & 5.5  \\
			&& $f_0$ & 0.073 & -0.78 & 1.6 \\
			&& $f_T$ & 0.032 & -0.45 & 1.6 \\
			\hline
			& \multirow{4}{*}{$B_c \to D^*$} & $V$ & 0.28 & -3.7 & 13 \\
			&& $A_0$ & 0.20 & -3.1 & 13   \\
			&& $A_1$ & 0.12 & -1.0 & 1.3  \\
			&& $A_2$ & 0.082 & -0.56 & -1.0 \\
			\hline
			& \multirow{4}{*}{$B_c \to D_s$} & $f_s$ & 0.65 & -6.5 & 13  \\
			&& $f_+$ & 0.11 & -1.6 & 6.2  \\
			&& $f_0$ & 0.087 & -0.85 & 1.6 \\
			&& $f_T$ & 0.037 & -0.50& 1.8 \\
			\hline
			& \multirow{4}{*}{$B_c \to D_s^*$} & $V$ & 0.29 & -3.8 & 13 \\
			&& $A_0$ & 0.21 & -3.2 & 13   \\
			&& $A_1$ & 0.13 & -1.1 & 1.2  \\
			&& $A_2$ & 0.095 & -0.61 & -1.3
			\label{parameter}
		\end{tabular}
	\end{ruledtabular}
\end{table}
\begin{figure}[htbp]
	\centering
	\includegraphics[width=8.5cm]{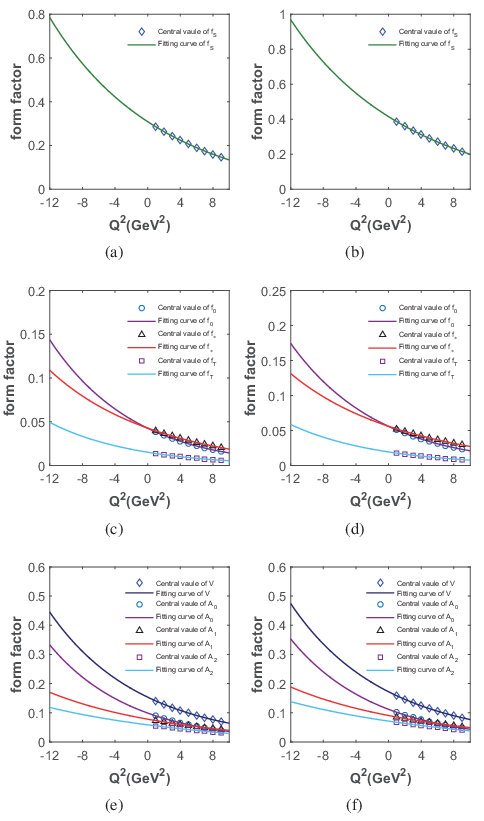}
	\caption{The fitting diagrams of form factors. The first column ((a), (c) and (e)) show the fitting results of form factors $B_c \to D^{(*)}$ transition, while those of $B_c \to D_s^{(*)}$ are shown in the second column ((b), (d) and (f)).}
	\label{FormFactor}
\end{figure}

After all of these parameters are determined, the form factors in the space-like region (1 GeV$^2 \leq Q^2 \leq$ 9 GeV$^2$) are calculated. Then these values are fitted into appropriate analytical function which is used to extrapolate the form factor into time-like region ($Q^2<0$). In this work, the $z-$series parameterizations approach is employed to realize this process.
For vector, axial vector and tensor form factors, the following parameterized function is adopted\cite{Bourrely:2008za, Cui:2022zwm},
\begin{align}\label{z}
	F({Q^2}) &= \frac{1}{{1 + {Q^2}/m_R^2}}\sum\limits_{k = 0}^{N - 1} {{b_k}[z{{({Q^2},{t_0})}^k} - {{( - 1)}^{k - N}}\frac{k}{N}z{{({Q^2},{t_0})}^N}]}
\end{align}
The scalar form factor is fitted in another form,
\begin{align}\label{z1}
	f_S({Q^2}) = \frac{1}{1 + {Q^2}/m_R^2}\sum\limits_{k = 0}^{N - 1} {{b_k}[z{{({Q^2},{t_0})}^k}]}
\end{align}
where $m_R$ is the mass of low-lying $B_c$ resonance\cite{Leljak:2019eyw}, and $z(Q^2,t_0)$ is written as,
\begin{align}
	z(Q^2,t_0) &= \frac{\sqrt {t_+ + Q^2} - \sqrt {t_+ - t_0}}{\sqrt {t_+ + Q^2}  + \sqrt {t_+ - t_0}}
\end{align}
Here, $t_0$ is a free parameter at the region (-$\infty$, $t_+$), it can be used to optimize the convergence of the series expansion. In present work the auxiliary parameter $t_0$ is taken as\cite{Bharucha:2010im, Cui:2022zwm,Leljak:2019eyw},
\begin{align}
	{t_0} &= {t_ + } - \sqrt {{t_ + }({t_ + } - {t_ - })}
\end{align}
where $t_\pm = (m_{B_c}\pm m_X)^2$. The expansion coefficients $b_i$ in Eqs. (\ref{z}) and (\ref{z1}) are obtained by fitting the numerical results with these two equations in space-like region.
All of the fitting parameters are listed in Tab. \ref{parameter} and the fitting diagrams are shown in Fig. \ref{FormFactor}. It can be seen that all of the numerical results are well fitted by the these fitting functions. Thus, we can obtain the values of form factors at $Q^2 = 0$ with these fitting functions. The results obtained in this work together with those of other collaborations are all summarized in Tab. \ref{compare}.
\begin{table*}[htbp]
	\centering
	\renewcommand\arraystretch{1.5}
	\begin{ruledtabular}
		\caption{Numerical results of the form factors $B_c \to D^{(*)}$ and $B_c \to D_{s}^{(*)}$ at $Q^2=0$.}
		\label{compare}
		\begin{tabular}{c c c c c c c }
			Modes & Form factors & This work & \cite{Zhang:2023ypl} & \cite{Dhir:2008zz} & \cite{Wang:2008xt} & \cite{Azizi:2008vy, Azizi:2008vv} \\
			\hline
			\multirow{4}{*}{$B_c \to D$} & $f_S$ & $0.31^{+0.10}_{-0.10}$ & - & - & - & -  \\
			& $f_+$ & $0.043^{+0.014}_{-0.014}$ & $0.17^{+0.01}_{-0.01}$ & 0.075 & 0.16 & 0.22  \\
			& $f_0$ & $0.043^{+0.014}_{-0.014}$ & $0.17^{+0.01}_{-0.01}$ & 0.075 & 0.16 & 0.22  \\
			& $f_T$ & $0.015^{+0.005}_{-0.005}$ & - & - & - & - \\
			\hline
			\multirow{4}{*}{$B_c \to D^*$} & $V$ & $0.15^{+0.04}_{-0.04}$ & $0.20^{+0.03}_{-0.03}$ & 0.16 & 0.13 & 0.63 \\
			& $A_0$ & $0.099^{+0.028}_{-0.027}$ & $0.14^{+0.02}_{-0.02}$ & 0.081 & 0.09 & 0.34  \\
			& $A_1$ & $0.078^{+0.022}_{-0.021}$ & $0.13^{+0.02}_{-0.02}$ & 0.095 & 0.08 & 0.41  \\
			& $A_2$ & $0.059^{+0.016}_{-0.016}$& $0.12^{+0.01}_{-0.01}$ & 0.11 & 0.07 & 0.45 \\
			\hline
			\multirow{4}{*}{$B_c \to D_s$} & $f_S$ & $0.41^{+0.12}_{-0.12}$ & - & - & - & - \\
			& $f_+$ & $0.056^{+0.016}_{-0.016}$ & $0.21^{+0.01}_{-0.01}$ & 0.15 & 0.28 & 0.16 \\
			& $f_0$ & $0.056^{+0.016}_{-0.016}$ & $0.21^{+0.01}_{-0.01}$ & 0.15 & 0.28 & 0.16 \\
			& $f_T$ & $0.020^{+0.006}_{-0.006}$ & - & - & - & -  \\
			\hline
			\multirow{4}{*}{$B_c \to D_s^*$} & $V$ & $0.17^{+0.04}_{-0.04}$ & $0.25^{+0}_{-0}$ & 0.29 & 0.23 & 0.54 \\
			& $A_0$ & $0.11^{+0.03}_{-0.03}$ & $0.18^{+0.02}_{-0.03}$ & 0.16 & 0.17 & 0.30  \\
			& $A_1$ & $0.091^{+0.023}_{-0.022}$ & $0.16^{+0.01}_{-0.02}$ & 0.18 & 0.14 & 0.36  \\
			& $A_2$ & $0.072^{+0.018}_{-0.017}$ & $0.15^{+0.01}_{-0.01}$ & 0.20 & 0.12 & 0.24 \\
		\end{tabular}
	\end{ruledtabular}
\end{table*}

It is shown that our results about $f_+$ and $f_0$ are roughly compatible with the results of BSW relativistic quark model\cite{Dhir:2008zz}, but too small compared to those of other methods. The possible reason for this divergence is the Coulomb-like $\alpha_s/v$ correction for the heavy quarkonium $B_c$, which in general can lead to increase of the results by two or three times\cite{Kiselev:2002vz}. For the form factors of $B_c\to D^{*}$ and $B_c\to D^{*}_{s}$, numerical results predicted in this work and those in Refs. \cite{Zhang:2023ypl, Dhir:2008zz, Wang:2008xt} are consistent with each other, while the values in Ref. \cite{Azizi:2008vv} are comparatively large. Although there are some differences between the numerical results obtained by different methods, these results exhibit similar characteristic. For example, the vector form factor $V$ is larger than axial vector form factor $A_{0,1,2}$. As for the form factors of $B_c\to D$ and $B_c\to D_{s}$, the value of $f_{s}$ is obviously larger than the others.

\section{Nonleptonic decays of $B_{c}$ meson}\label{sec4}
\begin{figure*}[htbp]
	\centering
	\includegraphics[width=14cm]{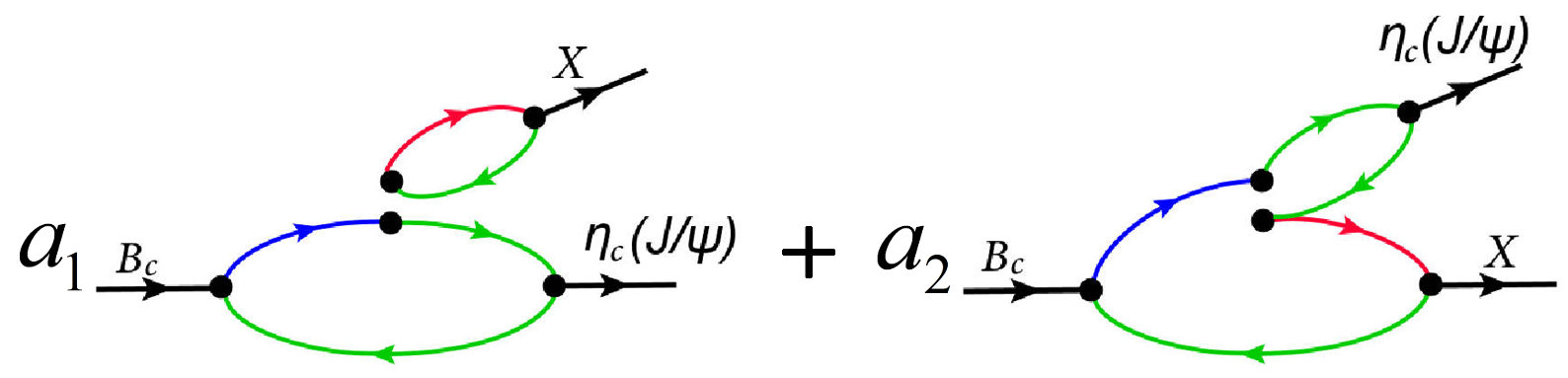}
	\caption{Feynman diagram of $B_c$ decaying to charmonium plus $X$ meson($X$=$D^{*}$, $D_s^{*}$,$D$, $D_s$). The solid lines with color of red, green, and blue stand for propagators of $d/s$, $c$ and $b$ quarks, respectively.}
	\label{decay2}
\end{figure*}
In our previous work, decay widths of several two-body decays for $B_c$ to charmonium plus one light meson were predicted\cite{Yu:2024utx}. As a continuation, two-body decay processes of $B_c$ to charmonium plus $D^{(*)}$ or $D_s^{(*)}$ mesons are analyzed in this work. This kind of decay process is realized according to weak decay of $b$-quark with $c$-quark acting as a spectator, which can be illustrated as Fig. \ref{decay2}. The effective Hamiltonian about this process has the following form,
\begin{align}
	H_{eff} = \frac{{G_F}}{{\sqrt 2 }}V_{cb}V_{cq}^*[{a_1}(\mu )O_1 + {a_2}(\mu )O_2]
\end{align}
where $G_F$ is the Fermi constant, $V_{cb}$ and $V_{cq}^*$ are the CKM matrix elements, $q$ denotes $d$ or $s$ quark, $a_1$ and $a_2$ are Wilson coefficients. $O_1$ and $O_2$ are four-fermion operators which are defined as,
\begin{align}
	\notag
	O_1 = \bar{c}\gamma_{\mu}(1-\gamma_5)b\bar{q}\gamma_{\mu}(1-\gamma_5)c \\
	O_2 = \bar{q}\gamma_{\mu}(1-\gamma_5)b\bar{c}\gamma_{\mu}(1-\gamma_5)c
\end{align}
The decay width of two-body decay process can be expressed as,
\begin{align}\label{decay1}
	\Gamma  = \frac{|\vec p|}{8\pi{m_{B_c}^2}} |T|^2
\end{align}
where $\vec p$ is the three-momentum of either of final particle in $B_c$ rest frame,
\begin{align}
	|\vec p| = \frac{\sqrt{\lambda(m_{B_c}^2, m_1^2, m_2^2)}}{2m_{B_c}}
\end{align}
In the framework of factorization approach, the matrix element $T$ in Eq. (\ref{decay1}) can be decomposed as the production of two matrix elements\cite{Yu:2022ngu},
\begin{align}\label{ME}
	\notag
	\left\langle {CX} \right|{O_1}\left| {{B_c}} \right\rangle  = \left\langle C \right|\bar c{\gamma _\mu }(1 - {\gamma _5})b\left| {{B_c}} \right\rangle \left\langle X \right|\bar q{\gamma _\mu }(1 - {\gamma _5})c\left| 0 \right\rangle \\
	\left\langle {CX} \right|{O_2}\left| {{B_c}} \right\rangle  = \left\langle X \right|\bar q{\gamma _\mu }(1 - {\gamma _5})b\left| {{B_c}} \right\rangle \left\langle C \right|\bar c{\gamma _\mu }(1 - {\gamma _5})c\left| 0 \right\rangle
\end{align}
with $X$=$D$, $D^{*}$, $D_s$, $D_s^{*}$ and $C$=$\eta_{c}$, $J/\psi$. The meson transition matrix elements at the right side of Eq. (\ref{ME}) have been parameterized as various form factors in Eq. (\ref{meson transition}), and meson vacuum matrix elements can be parameterized as the following decay constants,
\begin{align}\label{VM}
	\notag
	\left\langle P \right|\bar q{\gamma _\mu }(1-\gamma _5)q' \left| 0 \right\rangle &= if_Pq_\mu \\
	\left\langle V \right|\bar q{\gamma _\mu }(1-\gamma _5)q' \left| 0 \right\rangle &= f_Vm_V\xi_\mu
\end{align}
with $P$ denoting $D_{s}$, $D$ or $\eta_{c}$ meson and $V$ representing $D_{s}^{*}$, $D^{*}$ or $J/\psi$ meson. The values of form factors $B_{c}\to X$ have been determined in Sec. \ref{sec3} and those of $B_{c}\to C$ are taken from our previous work\cite{Yu:2024utx}. All of the values of parameters used in this section are also listed in Tab. \ref{parameters}.
With these above equations, the decay widths about these decay processes can be written as,
\begin{align}
	\notag
	&\Gamma(B_c \to D_{(s)} \eta_c) = \frac{|\vec{p}|}{16 \pi m_{B_c}^2} \Bigg\{G_F V_{cb} V_{cd(s)}^* \Big[a_{1}(m_{B_c}^2-m_{\eta_c}^2)f_{D_{(s)}} \\
	&\times f_0^{B_c \to \eta_c}(m_{D_{(s)}}^2) +a_{2}(m_{B_c}^2-m_{D_{(s)}}^2)f_{\eta_c}f_0^{B_c \to D_{(s)}}(m_{\eta_c}^2)\Big]\Bigg\}{^2}
\end{align}
\begin{align}
	\notag
	&\Gamma(B_c \to {D_{(s)}^*} {\eta_c}) = \frac{|\vec{p}|}{8\pi m_{B_c}^2}\Bigg\{G_F V_{cb} V_{cd(s)}^*\sqrt{\lambda(m_{B_c}^2,m_{\eta_c}^2,m_{D_{(s)}^*}^2)} \\
	&\times \Big[a_{1}f_{D_{(s)}^*}f_+^{B_c \to {\eta_c}}(m_{D_{(s)}^*}^2) + a_{2}f_{{\eta_c}}A_0^{B_c \to {D_{(s)}^*}}(m_{\eta_c}^2)\Big]\Bigg\}^2
\end{align}
\begin{align}
\notag
	&\Gamma(B_c \to {D_{(s)}} {J/\psi}) = \frac{|\vec{p}|}{8\pi m_{B_c}^2}\Bigg\{G_F V_{cb} V_{cd(s)}^*\sqrt{\lambda(m_{B_c}^2,m_{D_{(s)}}^2,m_{J/\psi}^2)} \\
	&\times \Big[a_{1}f_{{D_{(s)}}}A_0^{B_c \to {J/\psi}}(m_{D_{(s)}}^2)+a_{2}f_{J/\psi}f_+^{B_c \to {D_{(s)}}}(m_{J/\psi}^2)\Big]\Bigg\}^2
\end{align}
\begin{align}
	&\Gamma(B_c \to {D_{(s)}^*} {J/\psi}) = \frac{|\vec{p}|}{8\pi m_{B_c}^2}(|A_0|^2+|A_+|^2+|A_-|^2)
\end{align}
where $A_{0}, A_{+}, A_{-}$ are three polarization amplitudes, and can be written as,
\begin{widetext}
\begin{align}
	\notag
	&A_0(B_c \to D_{(s)}^*J/\psi) \\ \notag &
=\frac{G_F}{\sqrt{2}}V_{cb}V_{cd(s)}^*\Bigg\{ia_{1}f_{D_{(s)}^*}m_{D_{(s)}^*}\Big[-\frac{(m_{B_c}+m_{J/\psi})(m_{B_c}^2-m_{D_{(s)}^*}^2-m_{J/\psi}^2)}{2m_{D_{(s)}^*}m_{J/\psi}}
 A_1^{B_c \to J/\psi}(m_{D_{(s)}^*}^2)+\frac{\lambda(m_{B_c}^2,m_{D_{(s)}^*}^2,m_{J/\psi}^2)}{2m_{D_{(s)}^*}m_{J/\psi}(m_{B_c}+m_{J/\psi})}A_2^{B_c \to J/\psi}(m_{D_{(s)}^*}^2)\Big]\\ 	
	& +ia_{2}f_{J/\psi}m_{J/\psi}\Big[-\frac{(m_{B_c}+m_{D_{(s)}^*})(m_{B_c}^2-m_{D_{(s)}^*}^2-m_{J/\psi}^2)}{2m_{D_{(s)}^*}m_{J/\psi}} A_1^{B_c \to D_{(s)}^*}(m_{J/\psi}^2) +\frac{\lambda(m_{B_c}^2,m_{D_{(s)}^*}^2,m_{J/\psi}^2)}{2m_{D_{(s)}^*}m_{J/\psi}(m_{B_c}+m_{D_{(s)}^*})}A_2^{B_c \to D_{(s)}^*}(m_{J/\psi}^2)\Big]\Bigg\}
\end{align}
\begin{align}
	\notag
	&A_{\pm}(B_c \to {D_{(s)}^*}{J/\psi}) \\ \notag & =\frac{G_F}{\sqrt{2}}V_{cb}V_{cd(s)}^*\Bigg\{ia_{1}f_{D_{(s)}^*}m_{D_{(s)}^*}\Big[\pm\frac{\sqrt{\lambda(m_{B_c}^2,m_{D_{(s)}^*}^2,m_{J/\psi}^2)}}{m_{B_c}+m_{D_{(s)}^*}}V^{B_c \to {J/\psi}}(m_{D_{(s)}^*}^2)+(m_{B_c}+m_{D_{(s)}^*})A_1^{B_c \to {J/\psi}}(m_{D_{(s)}^*}^2)\Big]\\
	&+ia_{2}f_{J/\psi}m_{J/\psi}\Big[\pm\frac{\sqrt{\lambda(m_{B_c}^2,m_{D_{(s)}^*}^2,m_{J/\psi}^2)}}{m_{B_c}+m_{J/\psi}}V^{B_c \to {D_{(s)}^*}}(m_{J/\psi}^2)+(m_{B_c}+m_{J/\psi})A_1^{B_c \to {D_{(s)}^*}}(m_{J/\psi}^2)\Big]\Bigg\}
\end{align}
\begin{table*}
	\renewcommand\arraystretch{1.5}
	\begin{ruledtabular}
		\caption{Decay widths and Branching ratios of $B_c$ decaying to charmonium. Branching ratios are calculated at $\tau_{B_c}$ = 0.51 ps\cite{LHCb:2014ilr}.}\label{DBR}
		\begin{tabular}{c |c| c c c c c c}
			\multirow{2}{*}{Decay channels} & \multirow{2}{*}{Decay widths($10^{-7}$eV)} & \multicolumn{6}{c}{Branching ratios($10^{-3}$)}\\ \cline{3-8}
			&  &This work  & \cite{Zhang:2023ypl} & \cite{Kiselev:2002vz} & \cite{Ivanov:2006ni} & \cite{Rui:2014tpa} & \cite{Naimuddin:2012dy} \\
			\hline
			$B_c^- \to D \eta_c $ & $0.63^{+0.18}_{-0.16}$ & $0.048^{+0.014}_{-0.013}$ & $0.22^{0.03}_{-0.01}$ & 0.15 & 0.19 & $0.44^{+0.25}_{-0.18}$ & 0.06 \\
			$B_c^- \to D J/\psi $ & $0.46^{+0.13}_{-0.12}$ & $0.035^{+0.010}_{-0.009}$ & $0.20^{+0.03}_{-0.03}$ & 0.09 & 0.15 & $0.28^{+0.12}_{-0.08}$ & 0.04 \\
			$B_c^- \to D^* \eta_c $ & $0.86^{+0.25}_{-0.23}$ & $0.067^{+0.020}_{-0.018}$ & $0.31^{+0.02}_{-0.02}$ & 0.10 & 0.19 & $0.58^{+0.36}_{-0.25}$ & 0.07 \\
			$B_c^- \to D^* J/\psi $ & $2.2^{+0.6}_{-0.6}$ & $0.17^{+0.05}_{-0.04}$ & $0.41^{+0.06}_{-0.02}$ & 0.28 & 0.45 & $0.67^{+0.31}_{-0.19}$ & -\\
			$B_c^- \to D_s \eta_c $ & $16^{+5}_{-4}$ & $1.3^{+0.4}_{-0.3}$ & $6.44^{+1.78}_{-1.32}$ & 2.8 & 4.4 & $12.32^{+7.20}_{-5.56}$ & 1.79 \\
			$B_c^- \to D_s J/\psi $ & $11^{+3}_{-3}$ & $0.89^{+0.25}_{-0.23}$ & $6.09^{+1.62}_{-0.91}$ & 1.7 & 3.4 & $8.05^{+3.62}_{-2.08}$ & 1.15 \\
			$B_c^- \to D_s^* \eta_c $ & $21^{+6}_{-5}$ & $1.6^{+0.5}_{-0.4}$ & $6.97^{+0.68}_{-0.33}$ & 2.7 & 3.7 & $16.54^{+10.08}_{-8.74}$ & 1.49\\
			$B_c^- \to D_s^* J/\psi $ & $59^{+16}_{-15}$ & $4.6^{+1.2}_{-1.1}$ & $9.03^{+0.40}_{-0.38}$ & 6.7 & 9.7 & $20.45^{+10.24}_{-8.44}$ & -
		\end{tabular}
	\end{ruledtabular}
\end{table*}
\end{widetext}
The numerical results of the decay widths and branching ratios are all listed in Tab. \ref{DBR}.
It can be seen that the predicted branching ratios are smaller than most of the other predictions. This is mainly due to the smaller values of our predicted form factors, which will contribute to square reduction in the values of decay widths. Besides, it is also shown that the decay widths of $B_{c}\to D_{s}^{(*)}$ are larger than those of $B_{c}\to D^{(*)}$ processes. This is because the values of CKM matrix element $V_{cs}$ is much larger than $V_{cq}$(see Tab. \ref{parameters}) and the form factors also have the same characteristic. For example, the values of $f_{S}$ for $B_{c}\to D_{s}$ and $B_{c}\to D$ processes are $0.41^{+0.12}_{-0.12}$ and $0.31^{+0.10}_{-0.10}$, respectively (see Tab. \ref{compare}). Besides, the predicted branching ratios for most of the decay channels are in the range $10^{-5}$$\sim$$10^{-3}$ which lies in the detected ability of LHCb experiment. Thus all of these theoretical results can be verified by experiments in near future. Finally, using the results for decay process $B_c \to J/\psi \pi$ in our previous work\cite{Yu:2024utx}, the following ratios are obtained, $\frac{B(B_c \to J/\psi D_s)}{B(B_c \to J/\psi \pi)} = 3.3^{+2.5}_{-1.4}$, $\frac{B(B_c \to J/\psi D_s^*)}{B(B_c \to J/\psi D_s)} = 5.4^{+3.9}_{-2.3}$. These values are roughly consistent with the experimental data $2.90 \pm 0.57 \pm 0.24$ and $2.37 \pm 0.56 \pm 0.1$ \cite{LHCb:2013kwl}.

\begin{large}
\section{Conclusions}\label{sec5}
\end{large}
In this work, the form factors for $B_c \to D^{(*)}$ and $B_c \to D_s^{(*)}$ transition processes are systematically analyzed in the framework of three-point QCDSR. The numerical results in space-like region ($Q^2 > 0$) are firstly calculated and then are fitted into analytical functions using the $z-$series parameterizations approach. With these analytical functions, we obtain the form factors at $Q^2 =0$ and $Q^2 =- M^2$ by extrapolating the results into time-like region ($Q^2 < 0$). Using these form factors, the decay widths and branching ratios of two-body nonleptonic decays including $B_c \to \eta_c D^{*}$, $\eta_c D$, $ J/\psi D^{*}$, $ J/\psi D$, $\eta_c D_s^{*}$, $\eta_c D_s$, $J/\psi D_s^{*}$ and $J/\psi D_s$ are obtained with the factorization method. All of these results about form factors, decay widths and branching ratios obtained in this work not only provide useful information for further studying the heavy-quark dynamics but also are helpful to the experiments of heavy flavor physics in the future.

\section*{Acknowledgements}

This project is supported by National Natural Science Foundation, Grant Number 12175068 and Natural Science Foundation of HeBei Province, Grant Number A2018502124.


\begin{thebibliography}{}
	\bibitem{Gao:2010zzc}
	Y.~N.~Gao, J.~He, P.~Robbe, M.~H.~Schune and Z.~W.~Yang,
	Experimental prospects of the $B_c$ studies of the LHCb experiment,
	\href{https://doi.org/10.1088/0256-307X/27/6/061302}{Chin. Phys. Lett. \textbf{27}, 061302 (2010)}.
	
	\bibitem{Chen:2018obq}
	G.~Chen, C.~H.~Chang and X.~G.~Wu,
	$B_c (B_c^*)$ meson production via the proton-nucleus and the nucleus-nucleus collision modes at the colliders RHIC and LHC,
	\href{https://doi.org/10.1103/PhysRevD.97.114022}{Phys. Rev. D \textbf{97}, 114022 (2018)}.
	
	\bibitem{Gershtein:1994jw}
	S.~S.~Gershtein, V.~V.~Kiselev, A.~K.~Likhoded and A.~V.~Tkabladze,
	Physics of $B_c$ mesons,
	\href{https://doi.org/10.1070/PU1995v038n01ABEH000063}{Phys. Usp. \textbf{38}, 1 (1995)}.
	
	\bibitem{LHCb:2012ihf}
	R.~Aaij \textit{et al.} [LHCb],
	Measurements of $B_c^+$ production and mass with the $B_c^+ \to J/\psi \pi^+$ decay,
	\href{https://doi.org/10.1103/PhysRevLett.109.232001}{Phys. Rev. Lett. \textbf{109}, 232001 (2012)}.
	
	\bibitem{LHCb:2013xlg}
	R.~Aaij \textit{et al.} [LHCb],
	Observation of the Decay $B^+_c \to B^0_s \pi^+$,
	\href{https://doi.org/10.1103/PhysRevLett.111.181801}{Phys. Rev. Lett. \textbf{111}, 181801 (2013)}.
	
	\bibitem{LHCb:2013kwl}
	R.~Aaij \textit{et al.} [LHCb],
	Observation of $B^+_c \rightarrow J/\psi D_s^+$ and $B^+_c \rightarrow J/\psi D_s^{*+}$ decays,
	\href{https://doi.org/10.1103/PhysRevD.87.112012}{Phys. Rev. D \textbf{87}, 112012 (2013)}.
	
	\bibitem{Rui:2011qc}
	Z.~Rui, Z.~T.~Zou and C.~D.~Lu,
	The Two-Body $B_c {\to} D^{(*)}_{(s)}P$, $D^{(*)}_{(s)}V$ Decays in the Perturbative QCD Approach,
	\href{https://doi.org/10.1103/PhysRevD.86.074008}{Phys. Rev. D \textbf{86}, 074008 (2012)}.
	
	\bibitem{Rui:2012qq}
	Z.~Rui, Z.~Zhitian and C.~D.~Lu,
	The Double charm decays of $B_c$ Meson in the Perturbative QCD Approach,
	\href{https://doi.org/10.1103/PhysRevD.86.074019}{Phys. Rev. D \textbf{86}, 074019 (2012)}.
	
	\bibitem{Zou:2012sy}
	Z.~T.~Zou, X.~Yu and C.~D.~Lu,
	The $B_c\rightarrow D^{(*)}T$ decays in perturbative QCD approach,
	\href{https://doi.org/10.1103/PhysRevD.87.074027}{Phys. Rev. D \textbf{87}, 074027 (2013)}.
	
	\bibitem{Wang:2014yia}
	W.~F.~Wang, X.~Yu, C.~D.~Lu and Z.~J.~Xiao,
	Semileptonic decays $B_c^+ \to D_{(s)}^{(*)}{(l^+ \nu_l, l^+l^-, \nu \bar{\nu})}$ in the perturbative QCD approach,
	\href{https://doi.org/10.1103/PhysRevD.90.094018}{Phys. Rev. D \textbf{90}, 094018 (2014)}.
	
	\bibitem{Zhang:2024kjf}
	Z.~Q.~Zhang, Z.~Y.~Zhang, M.~X.~Xie, M.~Y.~Li and H.~X.~Guo,
	Branching ratios and CP asymmetries of the quasi-two-body decays $B_c \rightarrow \ K^{*}_0(1430,1950) D_{(s)} \rightarrow K \pi D_{(s)} $ in the PQCD approach,
	\href{https://doi.org/10.1140/epjc/s10052-024-13130-9}{Eur. Phys. J. C \textbf{84}, 864 (2024)}.
	
	\bibitem{Kiselev:2002vz}
	V.~V.~Kiselev,
	Exclusive decays and lifetime of $B_c$ meson in QCD sum rules,
	\href{https://arxiv.org/abs/hep-ph/0211021}{arXiv:hep-ph/0211021 [hep-ph] (2002).}
	
	\bibitem{Azizi:2007jx}
	K.~Azizi and V.~Bashiry,
	QCD sum rule analysis of the rare radiative $B_c \to D_{s}^* \gamma$ decay,
	\href{https://doi.org/10.1103/PhysRevD.76.114007}{Phys. Rev. D \textbf{76}, 114007 (2007)}.
	
	\bibitem{Azizi:2008vy}
	K.~Azizi and R.~Khosravi,
	Analysis of the rare semileptonic $B_c \to P(D, D_s ) \ell^{+} \ell^{-} / \nu \bar{\nu}$ decays within QCD sum rules,
	\href{https://doi.org/10.1103/PhysRevD.78.036005}{Phys. Rev. D \textbf{78}, 036005 (2008)}.
	
	\bibitem{Azizi:2008vv}
	K.~Azizi, F.~Falahati, V.~Bashiry and S.~M.~Zebarjad,
	Analysis of the Rare $B_c \to D^*_{s,d} l^+ l^-$ Decays in QCD,
	\href{https://doi.org/10.1103/PhysRevD.77.114024}{Phys. Rev. D \textbf{77}, 114024 (2008)}.
	
\bibitem{Wang:2024fwc}
Z.~G.~Wang,
$B_{c}$ meson and its scalar cousin with QCD sum rules,
	\href{https://doi:10.1088/1674-1137/ad5a71}{Chin. Phys. C \textbf{48}, 103104 (2024)}.


	\bibitem{Dhir:2008hh}
	R.~Dhir and R.~C.~Verma,
	$B_c$ Meson Form-factors and $B_c \to PV$ Decays Involving Flavor Dependence of Transverse Quark Momentum,
	\href{https://doi.org/10.1103/PhysRevD.79.034004}{Phys. Rev. D \textbf{79}, 034004 (2009)}.
	
	\bibitem{Dhir:2008zz}
	R.~Dhir, N.~Sharma and R.~C.~Verma,
	Flavor dependence of $B_c^+$ meson form factors and $B_c \to P P$ decays,
	\href{https://doi.org/10.1088/0954-3899/35/8/085002}{J. Phys. G \textbf{35}, 085002 (2008)}
	
	\bibitem{Wang:2008xt}
	W.~Wang, Y.~L.~Shen and C.~D.~Lu,
	Covariant Light-Front Approach for $B_c$ transition form factors,
	\href{https://doi.org/10.1103/PhysRevD.79.054012}{Phys. Rev. D \textbf{79}, 054012 (2009)}.
	
\bibitem{Chang:2019mmh}
Q.~Chang, X.~N.~Li and L.~T.~Wang,
Revisiting the form factors of $P\rightarrow V$ transition within the light-front quark models,
	\href{https://doi:10.1140/epjc/s10052-019-6949-3}{Eur. Phys. J. C \textbf{79}, 422 (2019)}.

	\bibitem{Li:2023wgq}
	X.~J.~Li, Y.~S.~Li, F.~L.~Wang and X.~Liu,
	Spectroscopic survey of higher-lying states of $B_c$ meson family,
	\href{https://doi.org/10.1140/epjc/s10052-023-12237-9}{Eur. Phys. J. C \textbf{83}, 1080 (2023)}.
	
	\bibitem{Zhang:2023ypl}
	Z.~Q.~Zhang, Z.~J.~Sun, Y.~C.~Zhao, Y.~Y.~Yang and Z.~Y.~Zhang,
	Covariant light-front approach for $B_c$ decays into charmonium: implications on form factors and branching ratios,
	\href{https://doi.org/10.1140/epjc/s10052-023-11576-x}{Eur. Phys. J. C \textbf{83}, 477 (2023)}.
	
	\bibitem{Li:2023mrj}
	Y.~S.~Li and X.~Liu,
	Angular distribution of the FCNC process $B_c \to D_s^*(\to D_s \pi)\ell^+\ell$,
	\href{https://doi.org/10.1103/PhysRevD.108.093005}{Phys. Rev. D \textbf{108}, 093005 (2023)}.
	
	\bibitem{Dubnicka:2017job}
	S.~Dubnicka, A.~Z.~Dubnickova, A.~Issadykov, M.~A.~Ivanov and A.~Liptaj,
	Study of $B_c$ decays into charmonia and $D$ mesons,
	\href{https://doi.org/10.1103/PhysRevD.96.076017}{Phys. Rev. D \textbf{96}, 076017 (2017)}.
	
	\bibitem{Ivanov:2002un}
	M.~A.~Ivanov, J.~G.~Korner and O.~N.~Pakhomova,
	The Nonleptonic decays $B^+_{c} \to D^+_{s} \bar{D}^0$ and $B^+_{c} \to D^+_{s} D^0$ in a relativistic quark model,
	\href{https://doi.org/10.1016/S0370-2693(03)00052-2}{Phys. Lett. B \textbf{555}, 189 (2003)}.
	
	\bibitem{Ivanov:2006ib}
	M.~A.~Ivanov, J.~G.~K\"orner and P.~Santorelli,
	Semileptonic and nonleptonic decays of $B_c$,
	\href{https://doi.org/10.1007/978-88-470-0530-3\_43}{arXiv:hep-ph/0609122[hep-ph] (2006)}.
	
	\bibitem{Huang:2008zg}
	T.~Huang, Z.~H.~Li, X.~G.~Wu and F.~Zuo,
	Semileptonic $B(B_s, B_c)$ decays in the light-cone QCD sum rules,
	\href{https://doi.org/10.1142/S0217751X0804189X}{Int. J. Mod. Phys. A \textbf{23}, 3237 (2008)}.
	
	\bibitem{Colangelo:1992cx}
	P.~Colangelo, G.~Nardulli and N.~Paver,
	QCD sum rules calculation of $B_c$ decays,
	\href{https://arxiv.org/10.1007/BF01555737}{Z. Phys. C \textbf{57}, 43 (1993)}.
	
	\bibitem{Wang:2007fs}
	Y.~M.~Wang and C.~D.~Lu,
	Weak productions of new charmonium in semi-leptonic decays of $B_c$,
	\href{https://doi.org/10.1103/PhysRevD.77.054003}{Phys. Rev. D \textbf{77}, 054003 (2008)}.
	
	\bibitem{Aliev:2010ac}
	T.~M.~Aliev, K.~Azizi and M.~Savci,
	Heavy $\chi_{Q_2}$ tensor mesons in QCD,
	\href{https://doi.org/10.1016/j.physletb.2010.05.018}{Phys. Lett. B \textbf{690}, 164 (2010)}.
	
	\bibitem{Wang:2012kw}
	Z.~G.~Wang,
	Analysis of the vector and axialvector $B_c$ mesons with QCD sum rules,
	\href{https://doi.org/10.1140/epja/i2013-13131-7}{Eur. Phys. J. A \textbf{49}, 131 (2013)}.
	
	\bibitem{Bashiry:2013waa}
	v.~Bashiry,
	Study of $\chi_{c0}(1P) \to J/\psi \gamma $ and $\chi_{b0}(1P) \to \Upsilon(1S) \gamma $ decays via QCD sum rules,
	\href{https://doi.org/10.1155/2014/432903}{Adv. High Energy Phys. \textbf{2014}, 432903 (2014)}.
	
	\bibitem{Peng:2019apl}
	Y.~Q.~Peng and M.~Z.~Yang,
	Form factors and decay of $\bar{B}_s^0\to J/\psi \phi$ from QCD sum rule,
	\href{https://doi.org/10.1142/S0217732320501874}{Mod. Phys. Lett. A \textbf{35}, 2050187 (2020)}.
	
	\bibitem{Shi:2019hbf}
	Y.~J.~Shi, W.~Wang and Z.~X.~Zhao,
	QCD Sum Rules Analysis of Weak Decays of Doubly-Heavy Baryons,
	\href{https://doi.org/10.1140/epjc/s10052-020-8096-2}{Eur. Phys. J. C \textbf{80}, 568 (2020)}.
	
	\bibitem{Lu:2023gmd}
	J.~Lu, G.~L.~Yu and Z.~G.~Wang,
	The strong vertices of charmed mesons $D$, $D^{*}$ and charmonia $J/\psi $, $\eta _{c}$,
	\href{https://doi.org/10.1140/epja/s10050-023-01115-3}{Eur. Phys. J. A \textbf{59}, 195 (2023)}.
	
	\bibitem{Lu:2023lvu}
	J.~Lu, G.~L.~Yu, Z.~G.~Wang and B.~Wu,
	Strong vertices of bottom mesons $B$ and $B^*$ and bottomonia $\Upsilon$, ${\eta}_b *$,
	\href{https://doi.org/10.1088/1674-1137/ad061d}{Chin. Phys. C \textbf{48}, 013102 (2024)}.
		
	\bibitem{Yu:2024utx}
	B.~Wu, G.~L.~Yu, J.~Lu and Z.~G.~Wang,
	Systematic analysis of the form factors of $B_c\rightarrow\eta_c$, $J/\psi$ and corresponding weak decays,
	\href{https://arxiv.org/abs/2406.08181}{arXiv:2406.08181 [hep-ph] (2024)}.
	
	\bibitem{Faustov:2019mqr}
	R.~N.~Faustov, V.~O.~Galkin and X.~W.~Kang,
	Semileptonic decays of $D$ and $D_s$ mesons in the relativistic quark model,
	\href{https://doi.org/10.1103/PhysRevD.101.013004}{Phys. Rev. D \textbf{101}, 013004 (2020)}.
	
	\bibitem{Wang:2014yza}
	Z.~G.~Wang and Z.~Y.~Di,
	Masses and decay constants of the heavy tensor mesons with QCD sum rules,
	\href{https://doi.org/10.1140/epja/i2014-14143-5}{Eur. Phys. J. A \textbf{50}, 143 (2014)}.
	
	\bibitem{Cutkosky:1960sp}
	R.~E.~Cutkosky,
	Singularities and discontinuities of Feynman amplitudes,
	\href{https://doi.org/10.1063/1.1703676}{J. Math. Phys. \textbf{1}, 429 (1960)}.
	
	\bibitem{Wang:2007ys}
	Y.~M.~Wang, H.~Zou, Z.~T.~Wei, X.~Q.~Li and C.~D.~Lu,
	The Transition form-factors for semi-leptonic weak decays of $J/\psi$ in QCD sum rules,
	\href{https://doi.org/10.1140/epjc/s10052-007-0498-x}{Eur. Phys. J. C \textbf{54}, 107 (2008)}.
			
	\bibitem{Bracco:2011pg}
	M.~E.~Bracco, M.~Chiapparini, F.~S.~Navarra and M.~Nielsen,
	Charm couplings and form factors in QCD sum rules,
	\href{https://doi.org/10.1016/j.ppnp.2012.03.002}{Prog. Part. Nucl. Phys. \textbf{67}, 1019 (2012)}
	
	\bibitem{ParticleDataGroup:2024cfk}
	S.~Navas \textit{et al.} [Particle Data Group],
	Review of particle physics,
	\href{https://doi.org/10.1103/PhysRevD.110.030001}{Phys. Rev. D \textbf{110}, 030001 (2024)}.
	
	\bibitem{Narison:2020guz}
	S.~Narison,
	QCD parameters, $f_{B_c}$ and $f_{B_c(2S)}$ from relativistic heavy quark sum rules,
	\href{https://doi.org/10.1016/j.nuclphysbps.2019.11.024}{Nucl. Part. Phys. Proc. \textbf{309}, 135 (2020)}.
	
	\bibitem{Wang:2015mxa}
	Z.~G.~Wang,
	Analysis of the masses and decay constants of the heavy-light mesons with QCD sum rules,
	\href{https://doi.org/10.1140/epjc/s10052-015-3653-9}{Eur. Phys. J. C \textbf{75}, 427 (2015)}.
	
	\bibitem{Narison:2010cg}
	S.~Narison,
	Gluon condensates and $c$, $b$ quark masses from quarkonia ratios of moments,
	\href{https://doi.org/10.1016/j.physletb.2011.09.116}{Phys. Lett. B \textbf{693}, 559 (2010)}.
	[erratum: Phys. Lett. B \textbf{705}, 544-544 (2011)]
	
	\bibitem{Narison:2011xe}
	S.~Narison,
	Gluon Condensates and precise $\overline{m}_{c,b}$ from QCD-Moments and their ratios to Order $\alpha_s^3$ and $\left\langle {G^4} \right\rangle $,
	\href{https://doi.org/10.1016/j.physletb.2011.11.058}{Phys. Lett. B \textbf{706}, 412 (2012)}.
	
	\bibitem{Narison:2011rn}
	S.~Narison,
	Gluon Condensates and $\bar{m}_b(\bar{m}_b)$ from QCD-Exponential Moments at Higher Orders,
	\href{https://doi.org/10.1016/j.physletb.2011.12.047}{Phys. Lett. B \textbf{707}, 259 (2012)}.
			
	\bibitem{Becirevic:2013bsa}
	D.~Be\v{c}irevi\'c, G.~Duplan\v{c}i\'c, B.~Klajn, B.~Meli\'c and F.~Sanfilippo,
	Lattice QCD and QCD sum rule determination of the decay constants of $\eta_c$, $J/\psi$ and $h_c$ states,
	\href{https://doi.org/10.1016/j.nuclphysb.2014.03.024}{Nucl. Phys. B \textbf{883}, 306 (2014)}.
		
	\bibitem{Cui:2022zwm}
	B.~Y.~Cui, Y.~K.~Huang, Y.~L.~Shen, C.~Wang and Y.~M.~Wang,
	Precision calculations of $B_{d,s} \to {\pi}$, $K$ decay form factors in soft-collinear effective theory,
	\href{https://doi.org/10.1007/JHEP03(2023)140}{JHEP \textbf{03}, 140 (2023)}.
	
	\bibitem{Bourrely:2008za}
	C.~Bourrely, I.~Caprini and L.~Lellouch,
	Model-independent description of $B \to \pi l \nu$ decays and a determination of $|V_{ub}|$,
	\href{https://doi.org/10.1103/PhysRevD.82.099902}{Phys. Rev. D \textbf{79}, 013008 (2009)}.
	
	\bibitem{Leljak:2019eyw}
	D.~Leljak, B.~Melic and M.~Patra,
	On lepton flavour universality in semileptonic $B_{c} \to \eta_{c}, J/{\psi}$ decays,
	\href{https://doi.org/10.1007/JHEP05(2019)094}{JHEP \textbf{05}, 094 (2019)}.

	\bibitem{Bharucha:2010im}
	A.~Bharucha, T.~Feldmann and M.~Wick,
	Theoretical and Phenomenological Constraints on Form Factors for Radiative and Semi-Leptonic B-Meson Decays,
	\href{https://doi.org/10.1007/JHEP09(2010)090}{JHEP \textbf{09}, 090 (2010)}.

	
	\bibitem{Yu:2022ngu}
	S.~Y.~Yu, X.~W.~Kang and V.~O.~Galkin,
	Two-body nonleptonic decays of the heavy mesons in the factorization approach,
	\href{https://doi.org/10.1007/s11467-023-1299-x}{Front. Phys. (Beijing) \textbf{18}, 64301 (2023)}.
	
	\bibitem{LHCb:2014ilr}
	R.~Aaij \textit{et al.} [LHCb],
	Measurement of the $B_c^+$ meson lifetime using $B_c^+ \to J\!/\!\psi \mu^+ \nu_{\mu} X$ decays,
	\href{https://doi.org/10.1140/epjc/s10052-014-2839-x}{Eur. Phys. J. C \textbf{74}, 2839 (2014)}.
	
	\bibitem{Ivanov:2006ni}
	M.~A.~Ivanov, J.~G.~Korner and P.~Santorelli,
	Exclusive semileptonic and nonleptonic decays of the $B_c$ meson,
	\href{https://doi.org/10.1103/PhysRevD.73.054024}{Phys. Rev. D \textbf{73}, 054024 (2006)}.
	
	\bibitem{Rui:2014tpa}
	Z.~Rui and Z.~T.~Zou,
	S-wave ground state charmonium decays of $B_c$ mesons in the perturbative QCD approach,
	\href{https://doi.org/10.1103/PhysRevD.90.114030}{Phys. Rev. D \textbf{90}, 114030 (2014)}.
	
	\bibitem{Naimuddin:2012dy}
	S.~Naimuddin, S.~Kar, M.~Priyadarsini, N.~Barik and P.~C.~Dash,
	Nonleptonic two-body $Bc$-meson decays,
	\href{https://doi.org/10.1103/PhysRevD.86.094028}{Phys. Rev. D \textbf{86}, 094028 (2012)}.

\end{thebibliography}
\end{document}